\documentclass[aps,pre]{revtex4}
\usepackage{mathrsfs}
\usepackage{amssymb}
\usepackage{graphicx}
\usepackage{amsmath}
\usepackage{graphicx}
\usepackage{dcolumn}
\usepackage{bm}

\newcommand{\brb}[1]{\left[#1\right]}
\newcommand{\brc}[1]{\left<#1\right>}
\newcommand{\bre}[1]{\left\{#1\right\}}

\newcommand{\be}{\begin{equation}}
\newcommand{\ee}{\end{equation}}
\newcommand{\bea}{\begin{eqnarray}}
\newcommand{\eea}{\end{eqnarray}}

\newcommand{\e}{\mathrm{e}}
\newcommand{\eps}{\epsilon}
\newcommand{\FBethe}{F_{\textrm{Bethe}}}

\newcommand{\cin}{c_ {\textrm{in}}}
\newcommand{\cout}{c_ {\textrm{out}}}

\newcommand{\argmax}{\arg\!\max}

\begin{document}
\title{Phase transitions in semisupervised clustering of sparse networks}
\author{
Pan Zhang$^1$,
Cristopher Moore$^1$,
and
Lenka Zdeborov\'a$^2$
}

\affiliation{
$^{1)}$Santa Fe Institute, Santa Fe, New Mexico 87501, USA\\
$^{2)}$Institut de Physique Th\'eorique, CEA Saclay and URA 2306, CNRS, Gif-sur-Yvette, France.}
\begin{abstract}
{Predicting labels of nodes in a network, such as community memberships or demographic variables, is an important problem with applications in social and biological networks. A recently-discovered phase transition puts fundamental limits on the accuracy of these predictions if we have access only to the network topology.  However, if we know the correct labels of some fraction $\alpha$ of the nodes, we can do better.  We study the phase diagram of this ``semisupervised'' learning problem for networks generated by the stochastic block model.  We use the cavity method and the associated belief propagation algorithm to study what accuracy can be achieved as a function of $\alpha$.  
For $k=2$ groups, we find that the detectability transition disappears for any $\alpha > 0$, in agreement with previous work. 
For larger $k$ where a hard but detectable regime exists, we find that the easy/hard transition (the point at which efficient algorithms can do better than chance) becomes a line of transitions where the accuracy jumps discontinuously at a critical value of $\alpha$.  
This line ends in a critical point with a second-order transition, beyond which the accuracy is a continuous function of $\alpha$.  
We demonstrate qualitatively similar transitions in two real-world networks.
}
\end{abstract}
\maketitle

\section{Introduction}
\label{sec:intro}

Community or module detection, also known as node clustering, is an important task in the study of biological, social and technological networks. Many methods have been proposed to solve this problem, including spectral clustering~\cite{Luxburg07,Newman2006,Krzakala2013}; modularity optimization~\cite{Newman2004,Newman2004fast,Clauset2004finding,Duch2005community}; statistical inference using generative models, such as the stochastic block model~\cite{Decelle2012asymptotic,Decelle2011inference,Karrer2011stochastic} and a wide variety of other methods, e.g.~\cite{Clauset2004finding,Blondel2008fast,Rosvall2008maps}.  See~\cite{Santo201075} for a review.  

It was shown in~\cite{Decelle2011inference,Decelle2012asymptotic} that for sparse networks generated by the stochastic block model~\cite{holland1983stochastic}, there is a phase transition in community detection.  This transition was initially established using the cavity method, or equivalently by analyzing the behavior of belief propagation. It was recently established rigorously in the case of two groups of equal size~\cite{mossel2012stochastic,massoulie2013community,mossel2013proof}.  In this case, below this transition, no algorithm can label the nodes better than chance, or even distinguish the network from an Erd\H{o}s-R\'enyi random graph with high probability.  In terms of belief propagation, there is a factorized fixed point where every node is equally likely to be in every group, and it becomes globally stable at the transition.

For more than two groups, there is an additional regime where the factorized fixed point is locally stable, but another, more accurate, fixed point is locally stable as well.  This regime lies between two spinodal transitions: the easy/hard transition where the factorized fixed point becomes locally unstable, so that efficient algorithms can achieve a high accuracy (also known as the Kesten-Stigum transition or the robust reconstruction threshold) and the transition where the accurate fixed point first appears (also known as the reconstruction threshold).  In between these two, there is a first order phase transition, where the Bethe free energy of these two fixed points cross.  This is the detectability transition, in the sense that an algorithm that can search exhaustively for fixed points---which would take exponential time---would choose the accurate fixed point above this transition.  However, below this transition there are exponentially many competing fixed points, each corresponding to a cluster of assignments, and even an exponential-time algorithm has no way to tell which is the correct one.  
(Note that, of these three transitions, the detectability transition is the only true thermodynamic phase transition; the others are dynamical.)  

In between the first order phase and easy/hard transitions, there is a ``hard but detectable'' regime where the communities can be identified in principle; if we could perform an exhaustive search, we would choose the accurate fixed point since it has lower free energy.  In Bayesian terms, the correct block model has larger total likelihood than an Erd\H{o}s-R\'enyi graph.  However, the accurate fixed point has a very small basin of attraction, making it exponentially hard to find---unless we have some additional information.

Here we model this additional information as a so-called
\emph{semisupervised} learning problem (e.g.~\cite{chapelle2006semi})
where we are given the true labels of some small fraction $\alpha$ of
the nodes.  This information shifts the location of these transitions;
in essence, it destabilizes the factorized fixed point, and pushes us 
towards the basin of attraction of the accurate one.  As a result, for some
values of the block model parameters, there is a discontinuous jump in
the accuracy as a function of $\alpha$.  Roughly speaking, for very
small $\alpha$ our information is local, consisting of the known nodes
and good guesses about nodes in their vicinity: but at a certain
$\alpha$ belief propagation causes this information to percolate,
giving us high accuracy throughout the network.  As we vary the block
model parameters, this line terminates at the point where the two spinodals 
and first order phase transitions all meet.  At that critical point there is a second-order phase transition, and beyond that point the accuracy is a continuous function of $\alpha$.

Semisupervised learning is an important task in machine learning, in settings where hidden variables or labels are expensive and time-consuming to obtain.  Semisupervised community detection was studied in several previous papers~\cite{allahverdyan2010community,steeg2013phase,eaton2012spin}. 
The conclusion of \cite{allahverdyan2010community} was that the detectability transition disappears for any $\alpha > 0$.  Later~\cite{steeg2013phase} suggested that in some cases it survives for more than two groups.  However, both these works were based on an approximate (zero temperature replica symmetric) calculation that corresponds to a far-from-optimal algorithm; moreover, it is known to lead to unphysical results in many other models such as graph coloring (which is a special case of the stochastic block model) or random $k$-SAT \cite{Zdeborova2007phase,mezard2002random}. 

In the present paper we investigate semisupervised community detection using the cavity method and belief propagation, which in sparse graphs is believed to be Bayes-optimal in the limit $n \to \infty$.  From a physics point of view, our results settle the question of what exactly happens in the semisupervised setting, including how the reconstruction, detectability, and easy/hard transitions vary as a function of $\alpha$.  From the point of view of mathematics, our calculations provide non-trivial conjectures that we hope will be amenable to rigorous proof. 

Our calculations follow the same methodology as those carried out in two other problems:
\begin{itemize}
 \item{\emph{Study of the ideal glass transition by random pinning.} 
 An important property of Bayes-optimal inference is that the true
     configuration cannot be distinguished from other
     configurations that are sampled at random from the
     posterior probability measure. This is why considering a
     disordered system similar to the one in this paper and fixing the
     value or position of a small fraction of nodes in a randomly
     chosen equilibrium configurations is formally the same problem to
     semisupervised learning. The analysis of systems with pinned
     particles was done in order to better understand the formation of
     glasses in~\cite{cammarota2012ideal,cammarota2013random}.
} 
 \item{\emph{Analysis of belief propagation guided decimation.}  belief propagation
      with decimation is a very interesting solver for a wide range of random
     constraint satisfaction problems. Its performance was analyzed in   
     \cite{Montanari2007solving,Ricci-Tersenghi2009on}.  If we decimate a fraction $\alpha$ of the variables (i.e., fix them to particular values) this affects the further performance of the algorithm in a way similar to semisupervised learning.}
\end{itemize}
For random $k$-SAT, a large part of this picture has been made rigorous~\cite{coja2011belief}.  Our hope is that similar techniques will apply to our results here.  As a first step, very recent work~\cite{Kanade14} shows that semisupervised learning does indeed allow for partial reconstruction below the detectability threshold for $k > 4$ groups.

The paper is organized as follows. Section \ref{sec:sbm} includes definitions and the description of the stochastic block model. In Sections~\ref{sec:synthetic} we consider semisupervised learning in the networks 
generated by stochastic block model. In Section~\ref{sec:real} we consider semisupervised learning in two real-world networks, finding transitions in the accuracy at a critical value of $\alpha$ qualitatively similar to our analysis for the block model.  We conclude in Section~\ref{sec:con}.

\section{The Stochastic Block Model, Belief Propagation, and Semisupervised Learning}
\label{sec:sbm}

The stochastic block model is defined as follows.  Nodes are split into $k$ groups, where each group $1 \le a \le k$ contains an expected 
fraction $q_a$ of the nodes.  Edge probabilities are given by a $k \times k$ matrix $p$.  We generate a random network $G$ with $n$ nodes as follows.  First, we choose a group assignment $t \in \{1,\ldots,k\}^n$ by assigning each node $i$ a label $t_i \in \{1,...,k\}$ chosen independently with probability $q_{t_i}$.  Between each pair of nodes $i$ and $j$, we then add an edge between them with probability $p_{t_i,t_j}$.  For now, we assume that the parameters $k$, $q$ (a vector denoting $\{q_a\}$), and $p$ (a matrix denoting $\{q_{ab}\}$) are known.

The likelihood of generating $G$ given the parameters and the labels is 
\begin{equation}
	P(G,t |q,p)= 
	\prod_{i} q_{t_i} 
	\prod_{\brc{ij}\in E} p_{t_i,t_j}
	\prod_{\brc{ij}\not\in E}(1-p_{t_i,t_j})  \, ,
\end{equation}
the Gibbs distribution of the labels $t$, i.e., their posterior distribution given $G$, can be computed via Bayes' rule,
\begin{equation}
	P(t | G,q,p ) = \frac{P(G,t|q,p) }{\sum_{s} P(G,s|q,p)} \, .
\end{equation}

In this paper, we consider sparse networks where $p_{ab} = c_{ab} / n$ for some constant matrix $c$.  In this case, the marginal probability $\psi^i_a$ that a given node $i$ has label $t_i=a$ can be computed using belief propagation.  The idea of BP is to replace these marginals with ``messages'' $\psi^{i \to \ell}_a$ from $i$ to each of its neighbors $\ell$, which are estimates of these marginals based on $i$'s interactions with its other neighbors~\cite{Yedidia2001understanding,Mezard2001the}.  We assume that the neighbors of each node are conditionally independent of each other; equivalently, we ignore the effect of loops.  For the stochastic block model, we obtain the following update equations for these messages~\cite{Decelle2011inference,Decelle2012asymptotic}:
\begin{equation}
\label{eq:update}
	\psi^{i \to \ell}_{a} = \frac{1}{Z_{i \to \ell}} \,q_{a} \e^{h_a} \prod_{j \in \partial i\backslash \ell}
	\sum_{b} c_{ab} \psi^{j\to i}_{b}  \, ,
\end{equation}
Here $Z_{i \to l}$ is a normalization factor and $h_a$ is an adaptive external field that enforces the expected group sizes, 
\begin{equation}
	h_a = \frac{1}{n} \sum_i \sum_b c_{ab} \psi_{b}^i,
\end{equation}
where the marginal probability that $t_i=a$ is given by
\begin{equation}
	\psi^{i}_a = \frac{1}{Z_{i}} \,q_a \e^{h_a} 
	\prod_{j \in \partial i} \sum_b c_{ab} \psi^{j\to i}_{b} \, ,
\end{equation}

In the usual setting, we start with random messages, and apply the BP equations~\eqref{eq:update} until we reach a fixed point.  In order to predict the node labels, we assign each node to its most-likely label according to its marginal:
\[
\widehat{t}_i = \argmax_a \psi^i_a \, . 
\]
Fixed points of the BP equations are stationary points of the Bethe free energy~\cite{Yedidia2001understanding}, which up to a constant is
\begin{equation}
	\FBethe = 
	\sum_{\brc{ij}\in E}\log\brb{\sum_{a,b=1}^kc_{ab}\psi_a^{i\to j}\psi_b^{j\to i}} - \sum_i\log\brb{\sum_{a=1}^k q_{a} \e^{h_{a}}\prod_{j\in \partial i}
	\sum_{\bre{b}}c_{ab} \psi^{j\to i}_{b} } \, .
\end{equation}
If there is more than one fixed point, the one with the lowest $\FBethe$ gives an optimal estimate of the marginals, since $\FBethe$ is minus the logarithm of the total likelihood of the block model.  
However, as we comment above, if the optimal fixed point has a small basin of attraction, then finding it through exhaustive search will take exponential time.  Analyzing stability of instability of these fixed points, including the trivial or ``factorized'' fixed point where $\psi^{i \to \ell}_a = q_a$, leads to the phase transitions in the stochastic block model described in~\cite{Decelle2011inference,Decelle2012asymptotic}. 

It is straightforward to adapt the above formalism to the case of semisupervised learning. One uses exactly the same equations except that nodes whose labels have been revealed have fixed messages.  If we know that $t_i = a^*$, then for all $\ell$ we have 
\[
\psi^{i \to \ell}_a = \delta_{a,a^*} \, . 
\]
Equivalently, we can define a local external field, replacing the global parameter $q$ with $q^i$ in~\eqref{eq:update}.  Then 
\[
q^i_a = \delta_{a,a^*} \, . 
\]

In this paper we focus on a widely-studied special case of the stochastic block model, also well-known as the planted partition model, where the groups are of equal size, i.e., ,$q_a=1/k$, and where $c_{ab}$ takes only two values:
\begin{equation}
	c_{ab} = 
	\begin{cases} \cin & \mbox{if $a=b$} \\ 
	\cout & \mbox{if $a\ne b$} \, . 
\end{cases}
\end{equation}
In that case, the average degree is $c = (\cin + (q-1) \cout)/q$.  It is common to parametrize this model with the ratio $\eps = \cout/\cin$.  When $\eps=0$ is small, nodes are connected only to others in the same group; at $\eps = 1$ the network is an Erd\H{o}s-R\'enyi graph, where every pair of nodes is equally likely to be connected.

We assume here that the parameters $k, \cin, \cout$ are known.  If they are unknown, inferring them from the graph and partial information about the nodes is an interesting learning problem in its own right.  One can estimate them from the set of known edges, i.e., those where both endpoints have known labels: in the sparse case there are $O(\alpha^2 n)$ such edges, or $O(n)$ if the fraction of known labels $\alpha$ is constant.  However, for $\alpha \sim 10^{-2}$, say, there are very few such edges until $n \gtrsim 10^5$ or $10^6$.  Alternately, we can learn the parameters using the expectation-maximization (EM) algorithm of~\cite{Decelle2011inference,Decelle2012asymptotic}, which minimizes the Bethe free energy, or a hybrid method where we initialize EM with parameters estimated from the known edges, if any.

\section{Results on the Stochastic Block Model and the Fate of the Transitions}
\label{sec:synthetic}

First we investigate semisupervised learning for assortative networks, i.e., the case $\cin > \cout$. As shown in \cite{Decelle2012asymptotic}, in the unsupervised case $\alpha=0$ there is a 
phase transition at 
\[
\cin- \cout = k \sqrt{c} \, , 
\]
where the factorized fixed point goes from stable to unstable.  Below this transition the overlap, i.e., the fraction of correctly assigned nodes, is $1/k$, no better than random chance.  For $k \le 4$ this phase transition is second-order: the overlap is continuous, but with discontinuous derivative at the transition.  For $k > 4$, it becomes an ``easy/hard'' transition, with a discontinuity in the overlap when we jump from the factorized fixed point to the accurate one.  In both cases, the convergence time (the number of iterations BP takes to reach a fixed point) diverges at the transition.

\begin{figure}
   \centering
    \includegraphics[width=0.45\columnwidth]{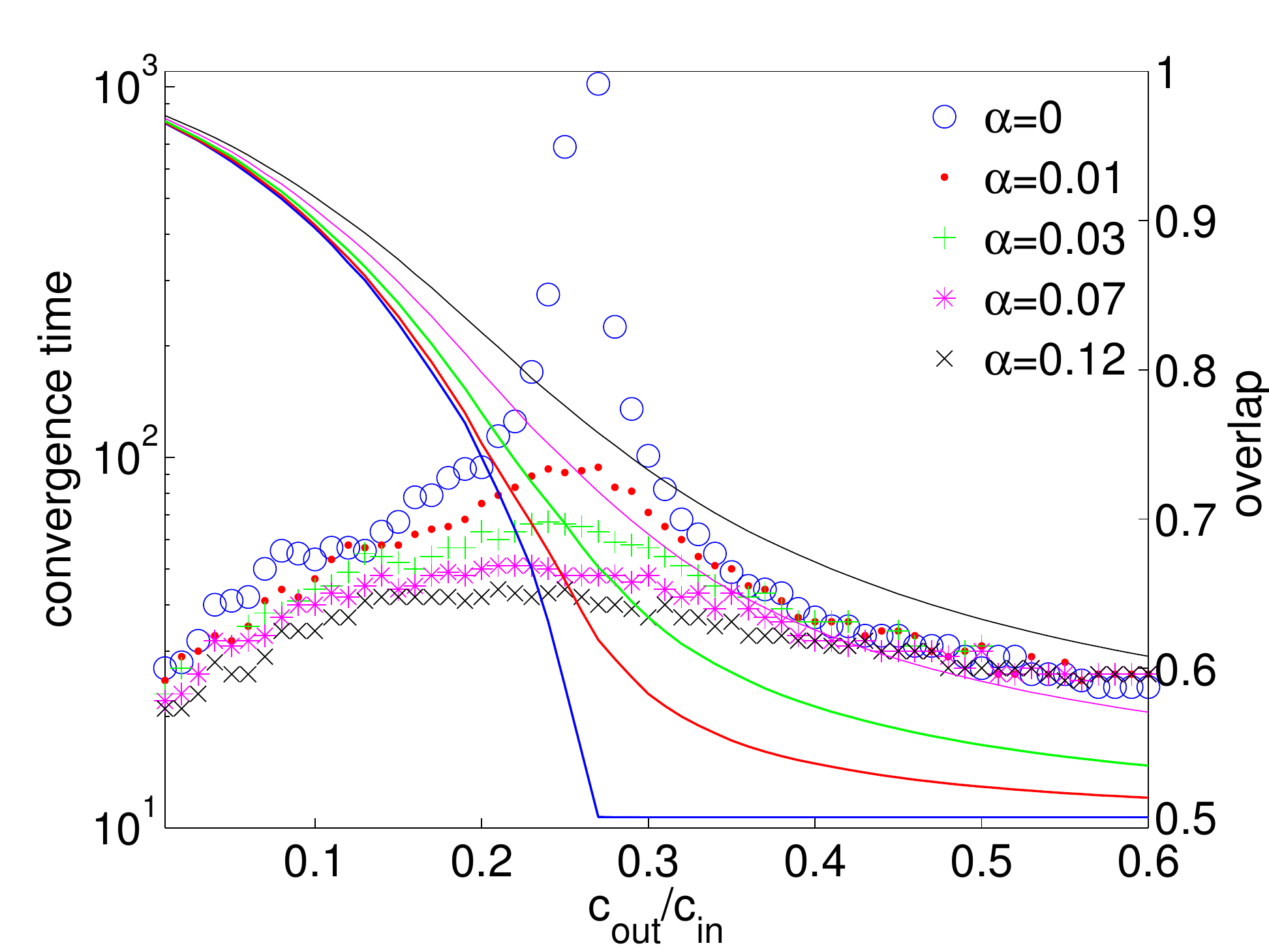} 
 \includegraphics[width=0.45\columnwidth]{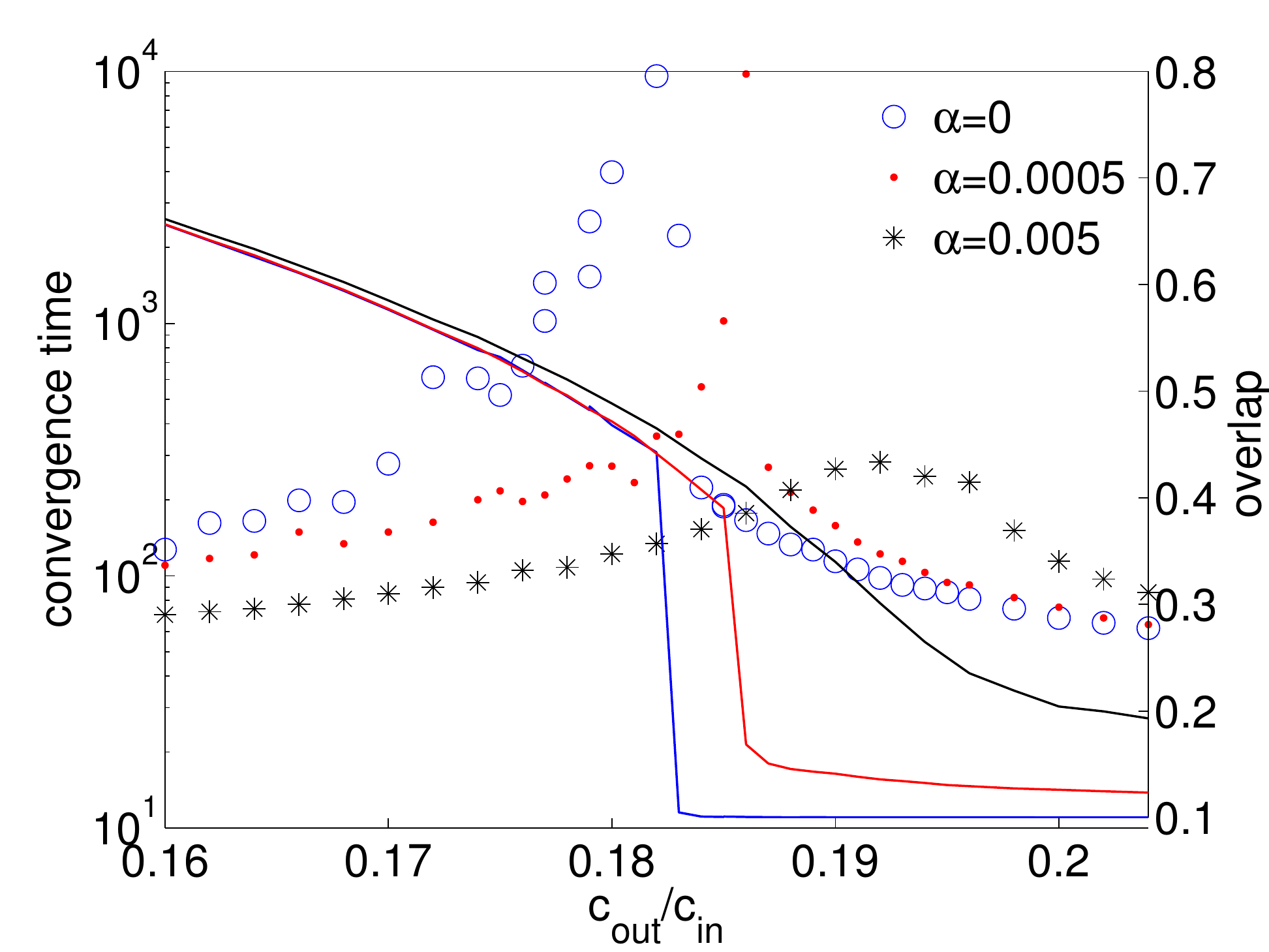} 

	\caption{Overlap and convergence time of BP as a function of $\epsilon=\cout/\cin$ for different $\alpha$, on networks generated by the stochastic block model.  On the left, $k=2$, $c=3$, and $n=10^5$.  For just two groups, the transition disappears for any $\alpha > 0$.  On the right, $k=10$, $c=10$, $n=5 \times 10^5$.  Here the easy/hard transition persists for small values of $\alpha$, with a discontinuity in the overlap and a diverging convergence time; this transition disappears at a critical value of $\alpha$.}
     \label{fig1}
\end{figure}

\begin{figure}
   \centering
    \includegraphics[width=0.45\columnwidth]{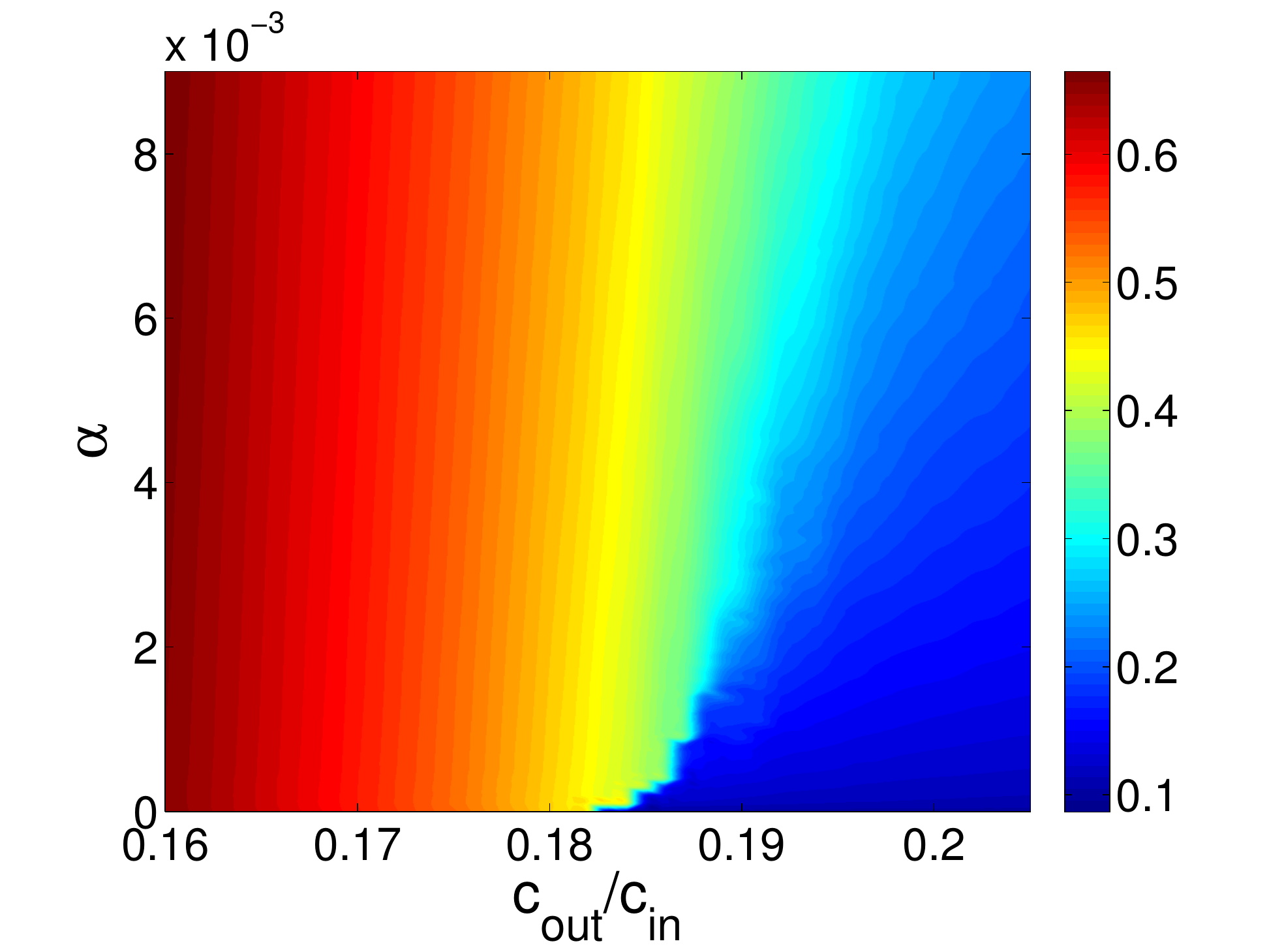} 
\includegraphics[width=0.45\columnwidth]{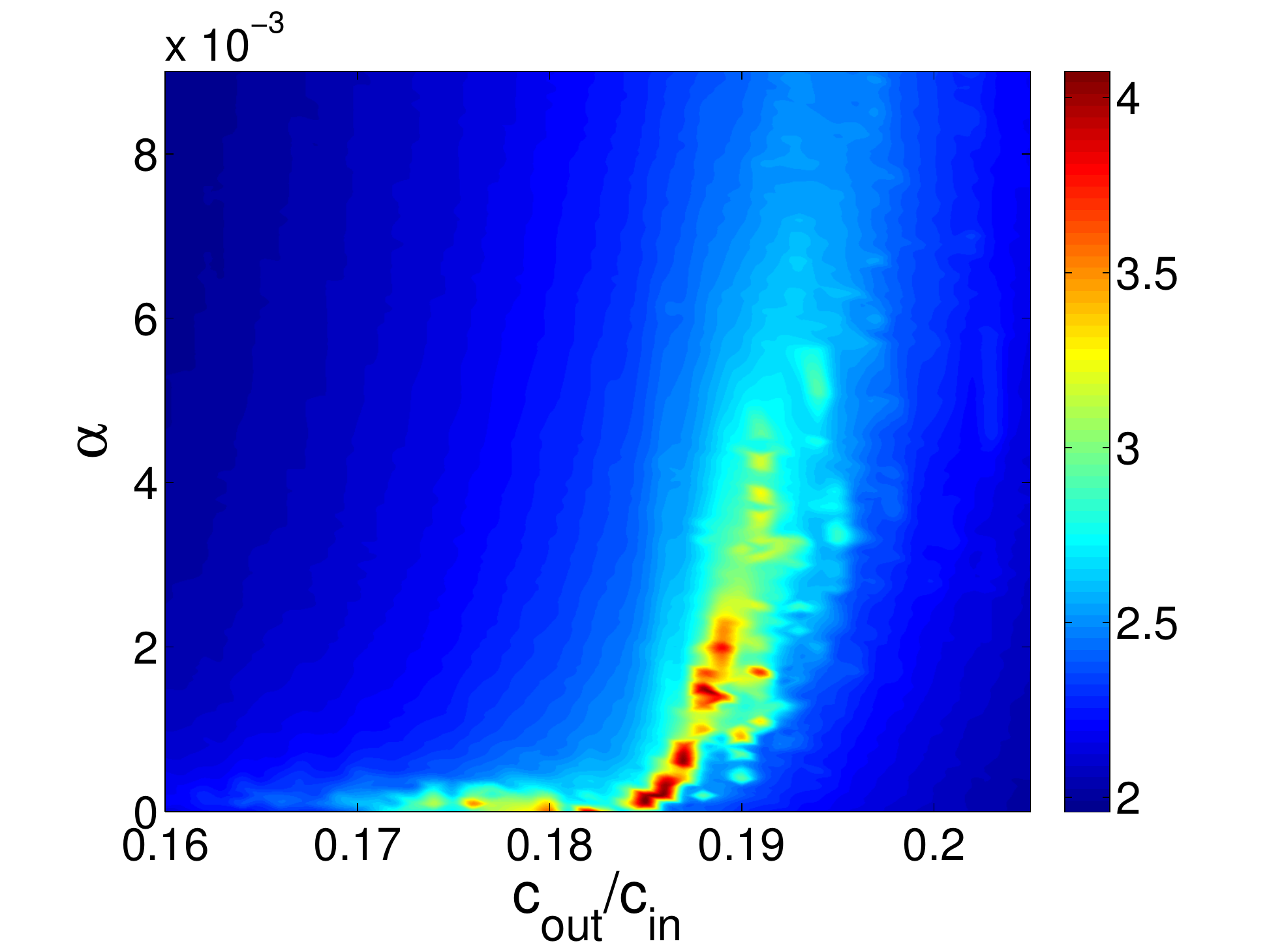} 
	\caption{Left, overlap as a function of $\epsilon=\cout/\cin$ and $\alpha$ for networks with same parameters as in the right of Fig.~\ref{fig1}.  The heat map shows a line of discontinuities, ending at a second-order phase transition beyond which the overlap is a smooth function.  Right, the logarithm (base 10) of the convergence time in the same plane, showing divergence along the critical line.}
    \label{fig2}
\end{figure}

In Fig.~\ref{fig1} we show the overlap achieved by BP for two different values of $k$ and various values of $\alpha$.  In each case, we hold the average degree $c$ fixed and vary $\eps = \cout/\cin$.  On the left, we have $k=2$.  Here, analogous to the unsupervised case $\alpha=0$, the overlap is a continuous function of $\eps$.  Moreover, for $\alpha > 0$ the detectability transition disappears: the overlap becomes a smooth function, and the convergence time no longer diverges.  This picture agrees qualitatively with the approximate analytical results in~\cite{allahverdyan2010community,steeg2013phase}. 


On the right-hand side of Fig.~\ref{fig1}, we show experimental results with $k=10$.  Here the easy/hard transition persists for sufficiently small $\alpha$, with a discontinuity in the overlap and a diverging convergence time.  At a critical value of $\alpha$, the transition disappears, and the overlap becomes a smooth function of $\eps$; beyond that point the convergence time has a smooth peak but does not diverge.  Thus there is a line of discontinuities, ending in a second-order phase transition at a critical point.  We show this line in the $(\alpha,\eps)$-plane in Fig.~\ref{fig2}.  On the left, we see the discontinuity in the overlap, and on the right we see that the convergence time diverges along this line.


Note that the authors of~\cite{steeg2013phase} also predicted the survival of the easy/hard discontinuity in the assortative case.  Their approximate computation, however, overestimates the strength of the phase transition, and misplaces
its position.  In particular, it predicts the discontinuity for all $k>2$, whereas it holds only for $k>4$.


The full physical picture of what happens to the ``hard but detectable'' regime in the semisupervised case, and to the spinodal and detectability transitions that define it, is very interesting.  To explain it in detail we focus on the disassortative case, and specifically the case of planted graph coloring where $\cin=0$.  The situation for the assortative case is qualitatively similar, but for graph coloring the discontinuity in the overlap is very strong and appears for any $k > 3$, making these phenomena easier to see numerically.

\begin{figure}
   \centering
\includegraphics[width=0.45\columnwidth]{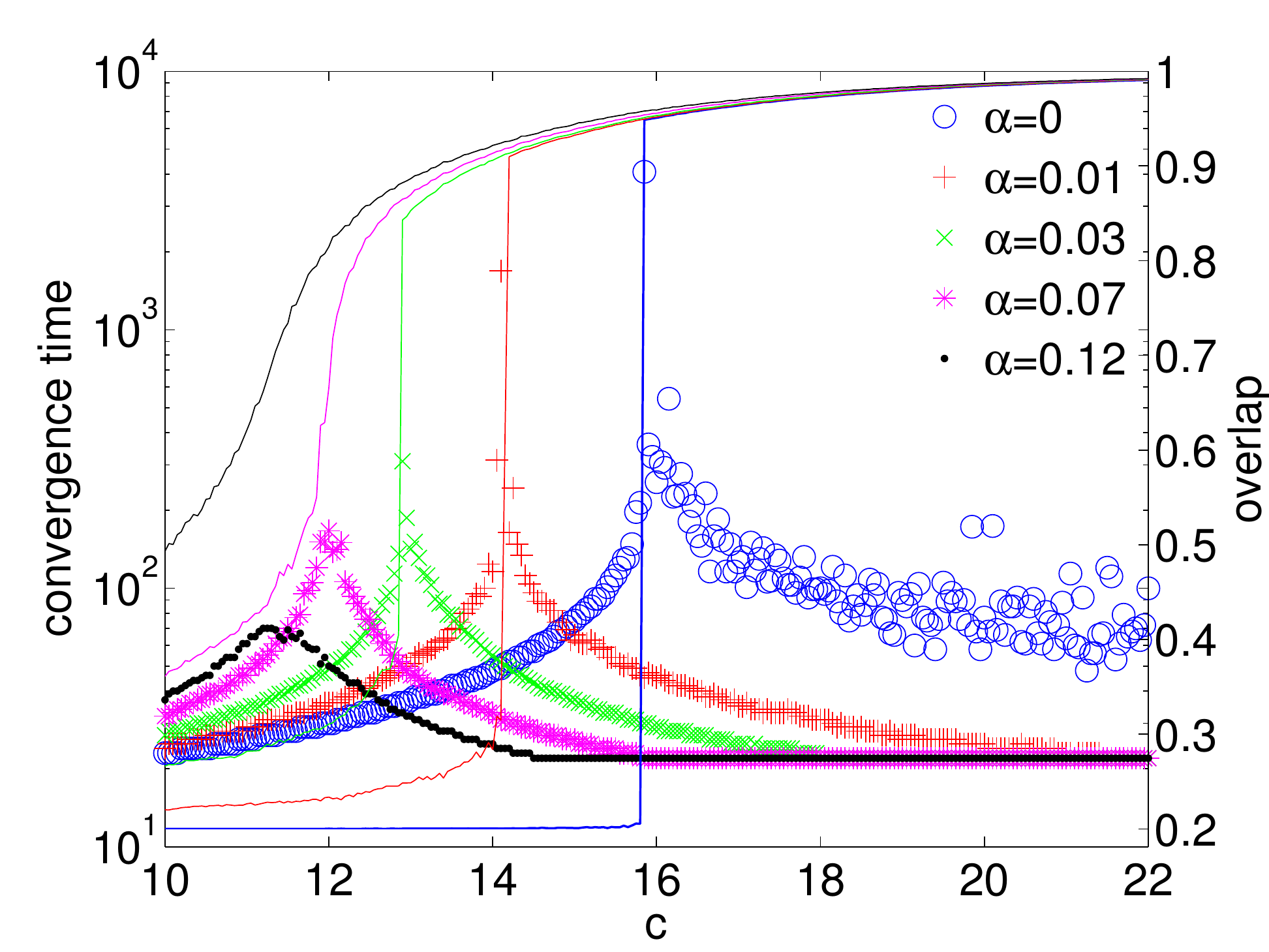} 
\includegraphics[width=0.45\columnwidth]{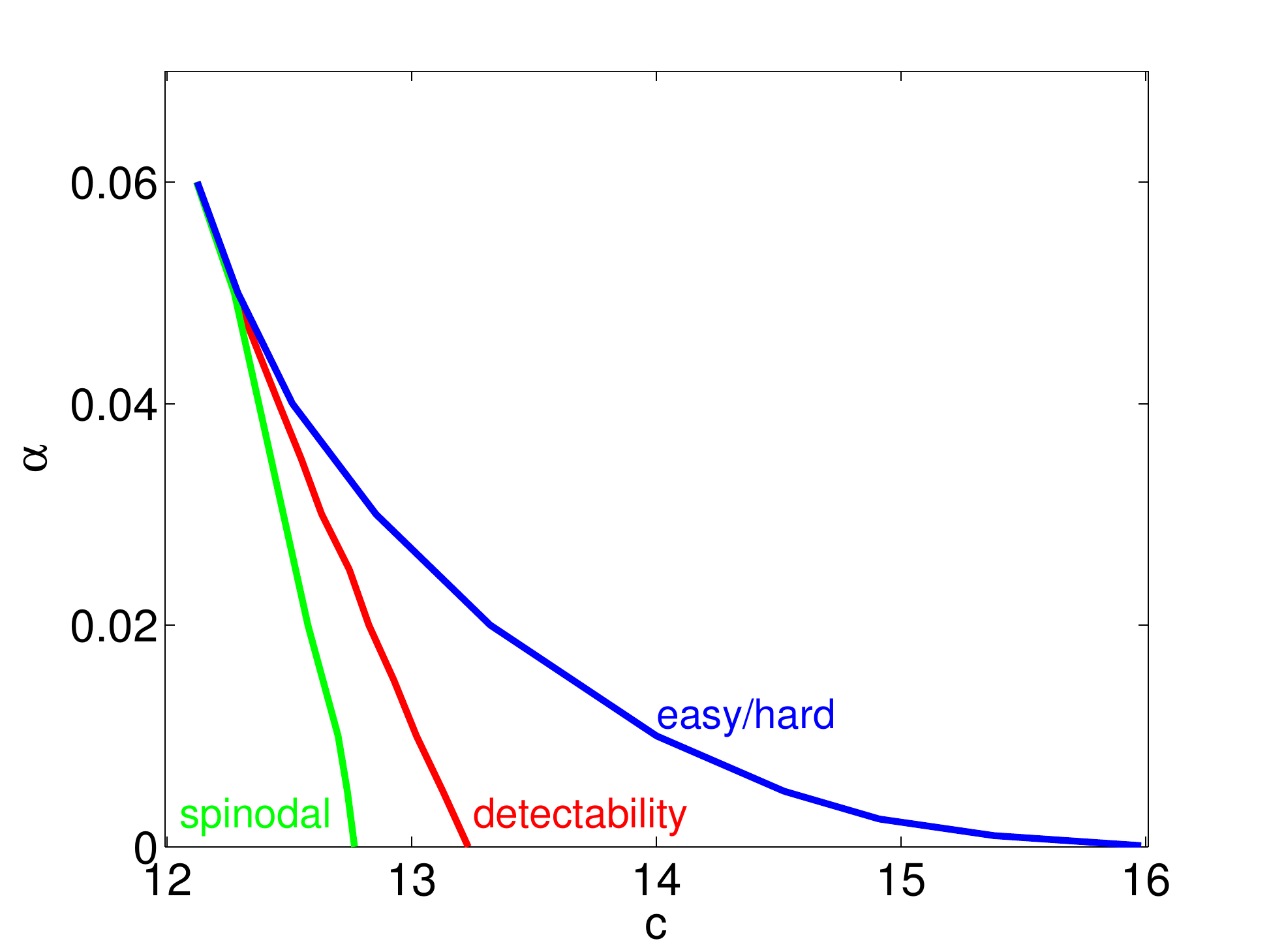} 

	\caption{Left, overlap and BP convergence time in the planted 5-coloring problem as a function of the average degree $c$ for various values of $\alpha$.  Right, the three transitions described in the text: the easy/hard or Kesten-Stigum transition where the factorized fixed point becomes stable (blue), the lower spinodal transition where the accurate fixed point disappears (green), and the detectability transition where the Bethe free energies of these fixed points cross (red).  The hard but detectable regime, where BP with random initial messages does no better than chance but exhaustive search would succeed, is between the red and blue lines.  All three transitions meet at a critical point, beyond which the overlap is a smooth function of $c$ and $\alpha$.  Here $n=10^5$.
}
\label{fig3}
\end{figure}

Fig.~\ref{fig3}, on the left, shows the overlap and convergence time of BP for $k=5$ colors.  In the unsupervised case $\alpha=0$, there are a total of three transitions as we decrease $c$ (making the problem of recovering the planted coloring harder).  The overlap jumps at the easy/hard spinodal transition, where the factorized fixed point becomes stable: this occurs at $c=(k-1)^2$.  At the lower spinodal transition, the accurate fixed point disappears. In between these two spinodal transitions, both fixed points exist.  Their Bethe free energies cross at the detectability transition: below this point, even a Bayesian algorithm with the luxury of exhaustive search would do no better than chance.  Thus the ``hard but detectable'' regime lies in between the detectability and easy/hard transitions~\cite{Zdeborova2007phase}.



On the right of Fig.~\ref{fig3}, we plot the two spinodal transitions, and the detectability transition in between them, in the $(c,\alpha)$-plane.  We see that these transitions persist up to a critical value of $\alpha \approx 0.06$.  At that point, all three meet at a second-order phase transition, beyond which the overlap is a smooth function.  The very same picture arises in the two related problems mentioned in the introduction, namely the glass transition with random pinning and BP-guided decimation in random $k$-SAT; see e.g. Fig.~1 in~\cite{cammarota2012ideal,cammarota2013random} and Fig.~3 in~\cite{Ricci-Tersenghi2009on}.  

\begin{figure}
   \centering
    \includegraphics[width=0.48\columnwidth]{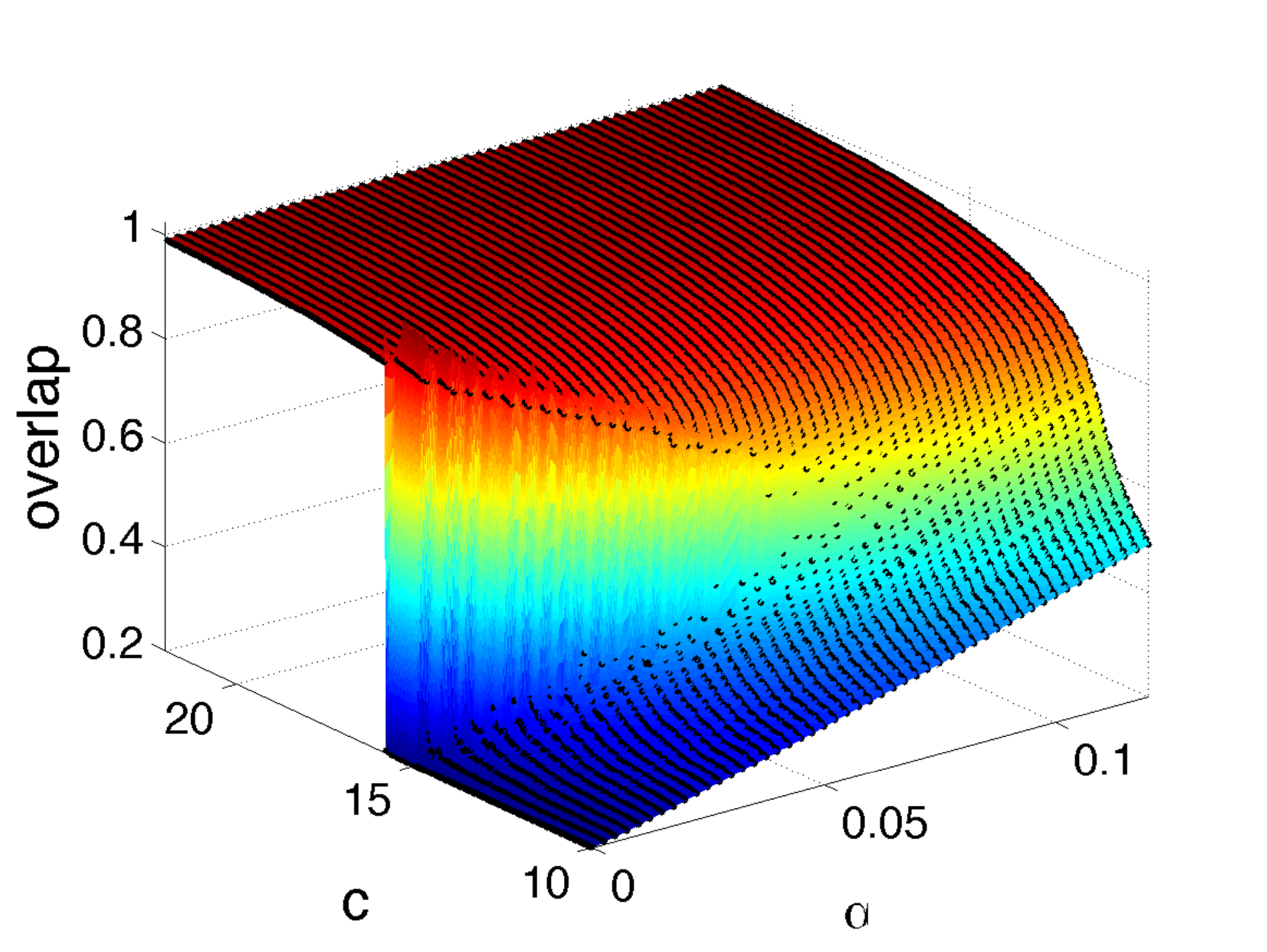}  \\
    \includegraphics[width=0.48\columnwidth]{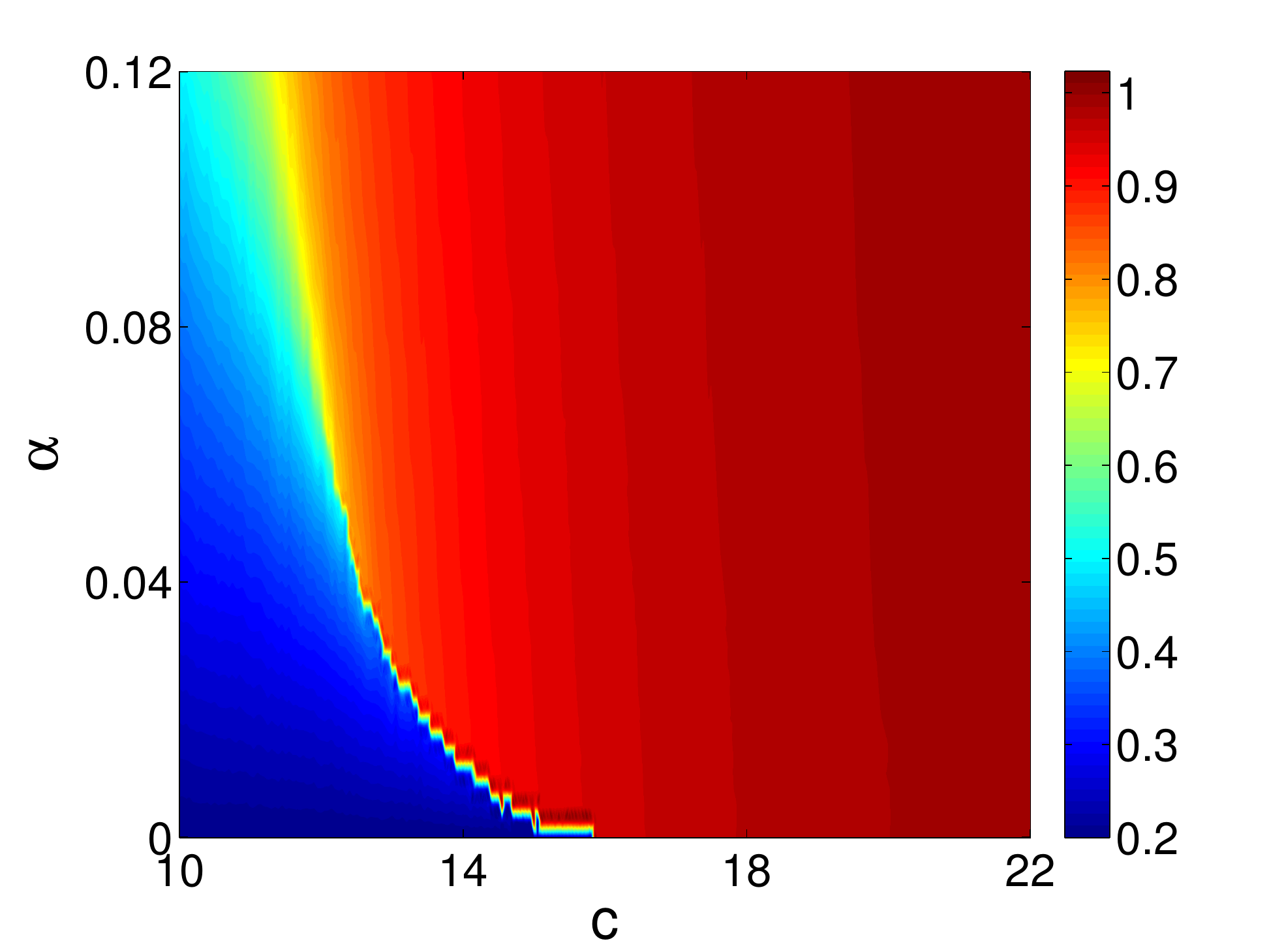} 
    \includegraphics[width=0.48\columnwidth]{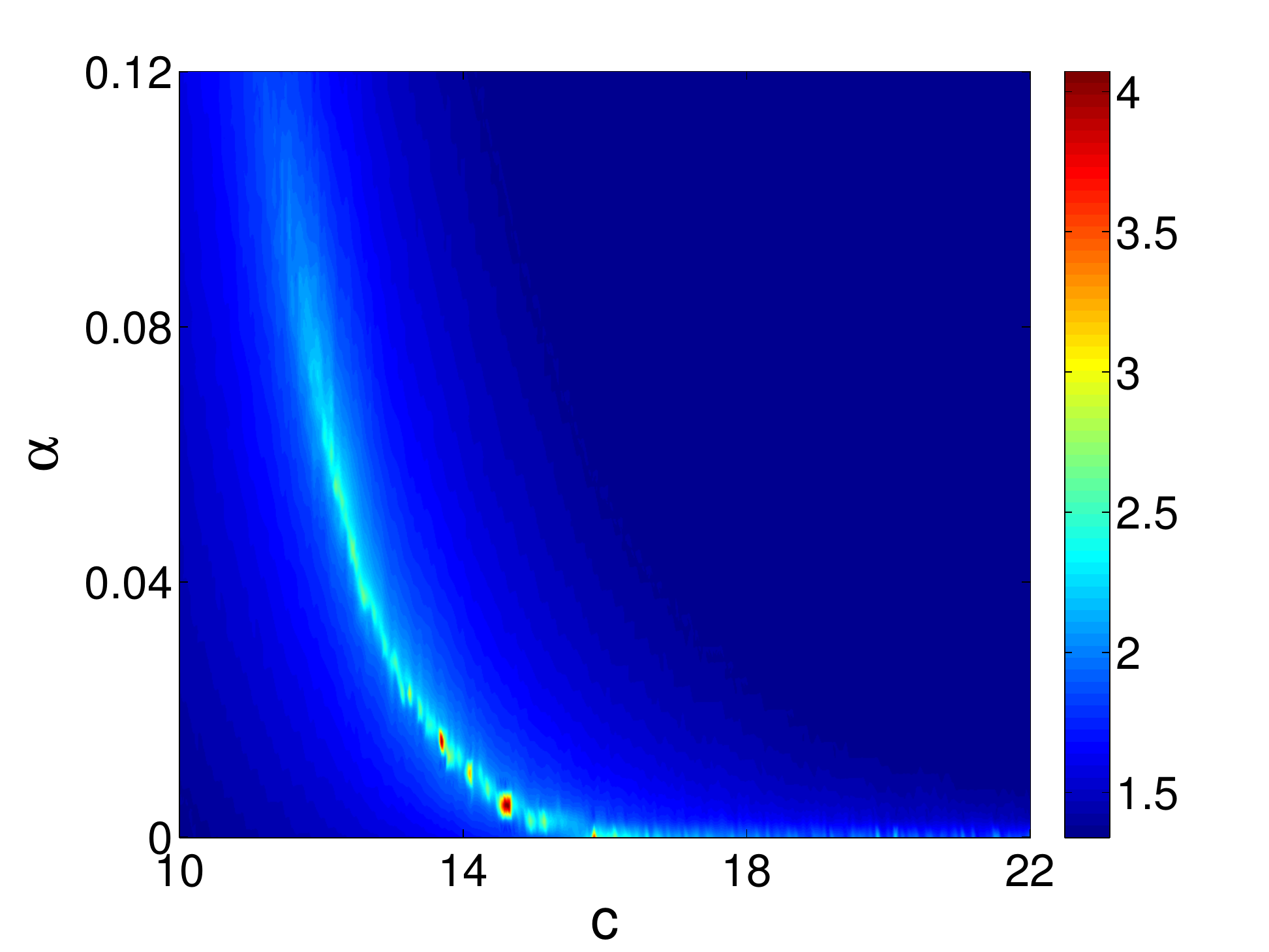} 
\caption{Overlap (top and left) and convergence time (right) as a function of the average degree $c$ and the fraction of known labels $\alpha$ for the planted $5$-coloring problem on networks with $n=10^5$.  The height of the discontinuity decreases until we reach the critical point. The convergence time diverges along the discontinuity.  Compare Fig.~\ref{fig2} for the assortative case.}
\label{fig4}
\end{figure}

Finally, in Fig.~\ref{fig4} we plot the overlap and convergence time for the planted $5$-coloring problem in the $(c,\alpha)$-plane.  Analogous to the assortative case in Fig.~\ref{fig2}, but more visibly, there is a line of discontinuities in the overlap along which the convergence time diverges; the height of the discontinuity decreases until we reach the critical point.

\section{Results on Real-World Networks}
\label{sec:real}

In this section we study semisupervised learning in real-world networks.  
Real-world networks are of course not generated by the stochastic block model; however, the block model can often achieve high accuracy for community detection, if its parameters are fit to the network.

To explore the effect of semisupervised learning, we set the parameters $q_a$ and $c_{ab}$ in two different ways.  In the first way, the algorithm is given the best possible values of these parameters in advance, as determined by the ground truth labels: this is cheating, but it separates the effect of being given $\alpha n$ node labels from the process of learning the parameters.  In the second (more realistic) way, the algorithm uses the expectation-maximization (EM) algorithm of~\cite{Decelle2011inference,Decelle2012asymptotic}, which minimizes the free energy.  As discussed in Section~\ref{sec:sbm}, we initialize the EM algorithm with parameters estimated from edges where both endpoints have known labels, if any.

\begin{figure}
   \centering
    \includegraphics[width=0.45\columnwidth]{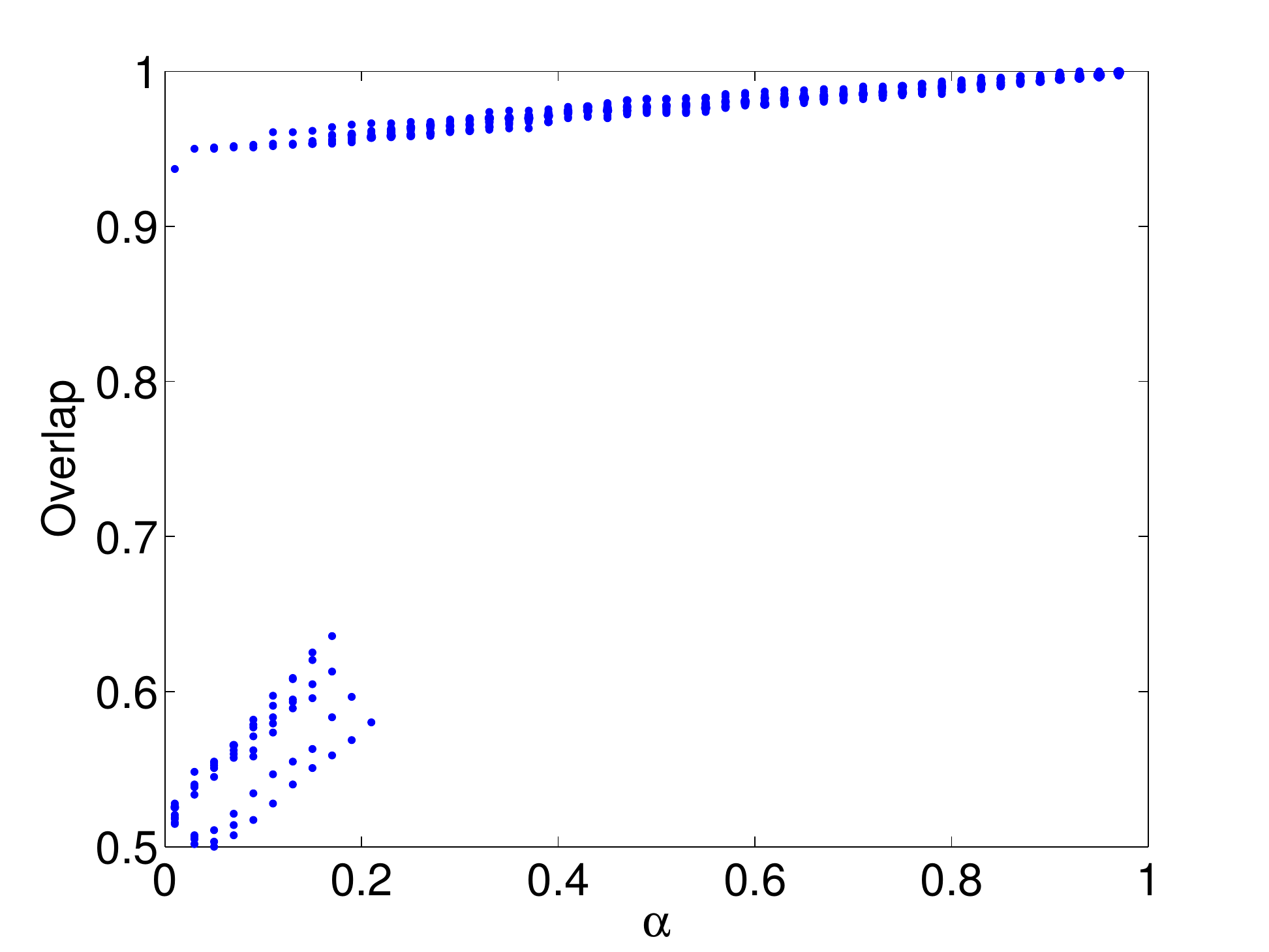} 
    \includegraphics[width=0.45\columnwidth]{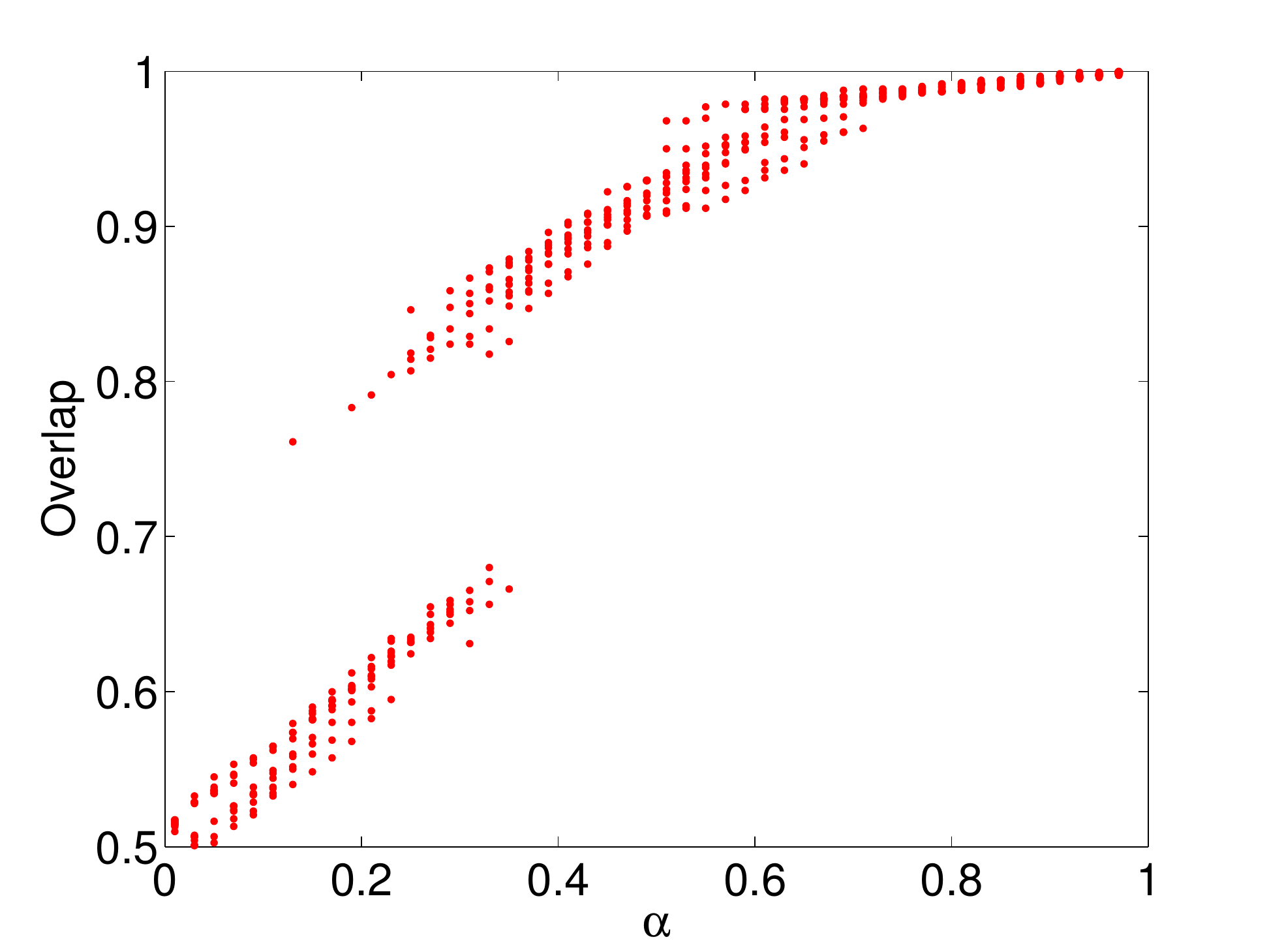} 
    \includegraphics[width=0.45\columnwidth]{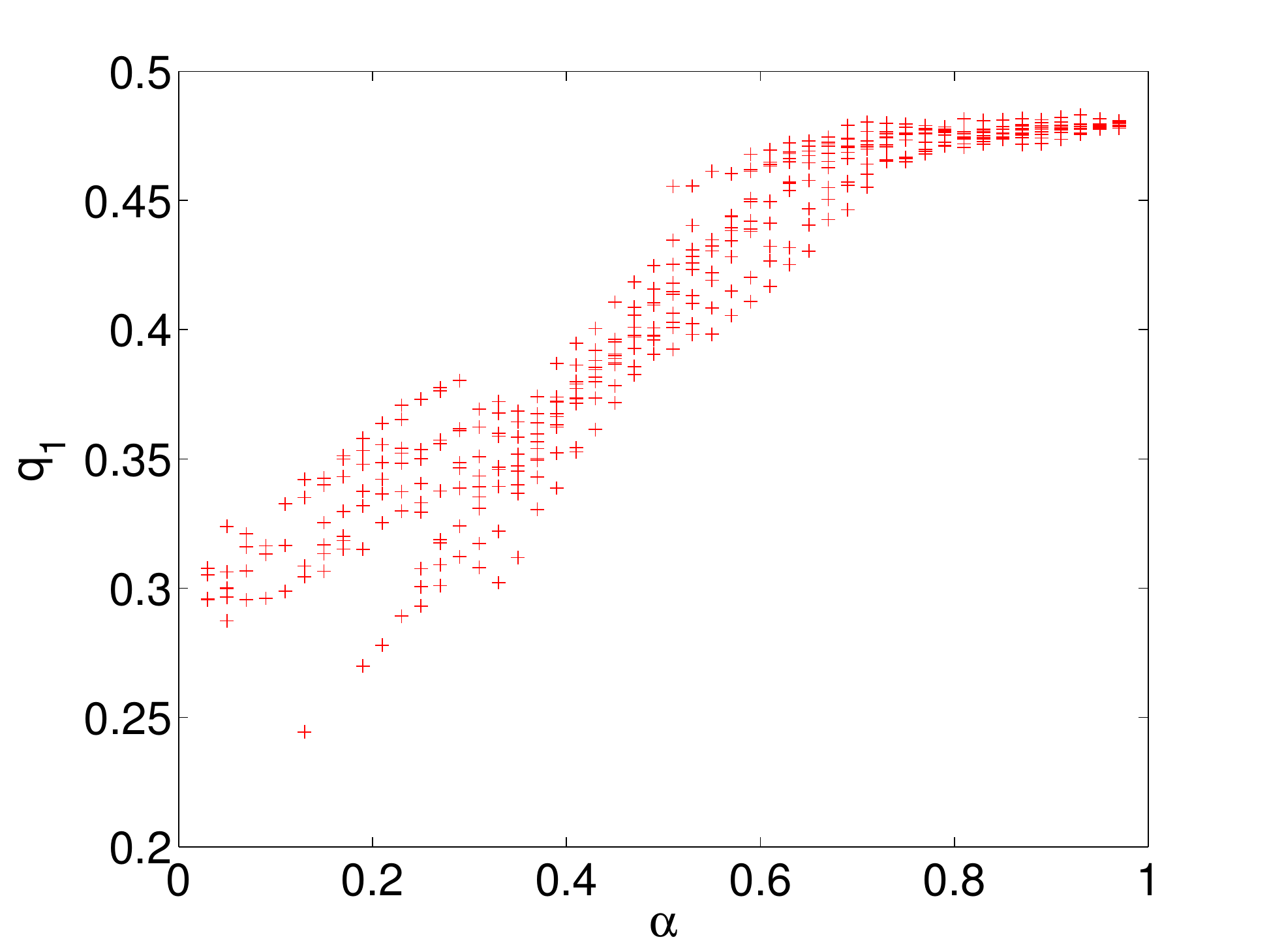} 
    \includegraphics[width=0.45\columnwidth]{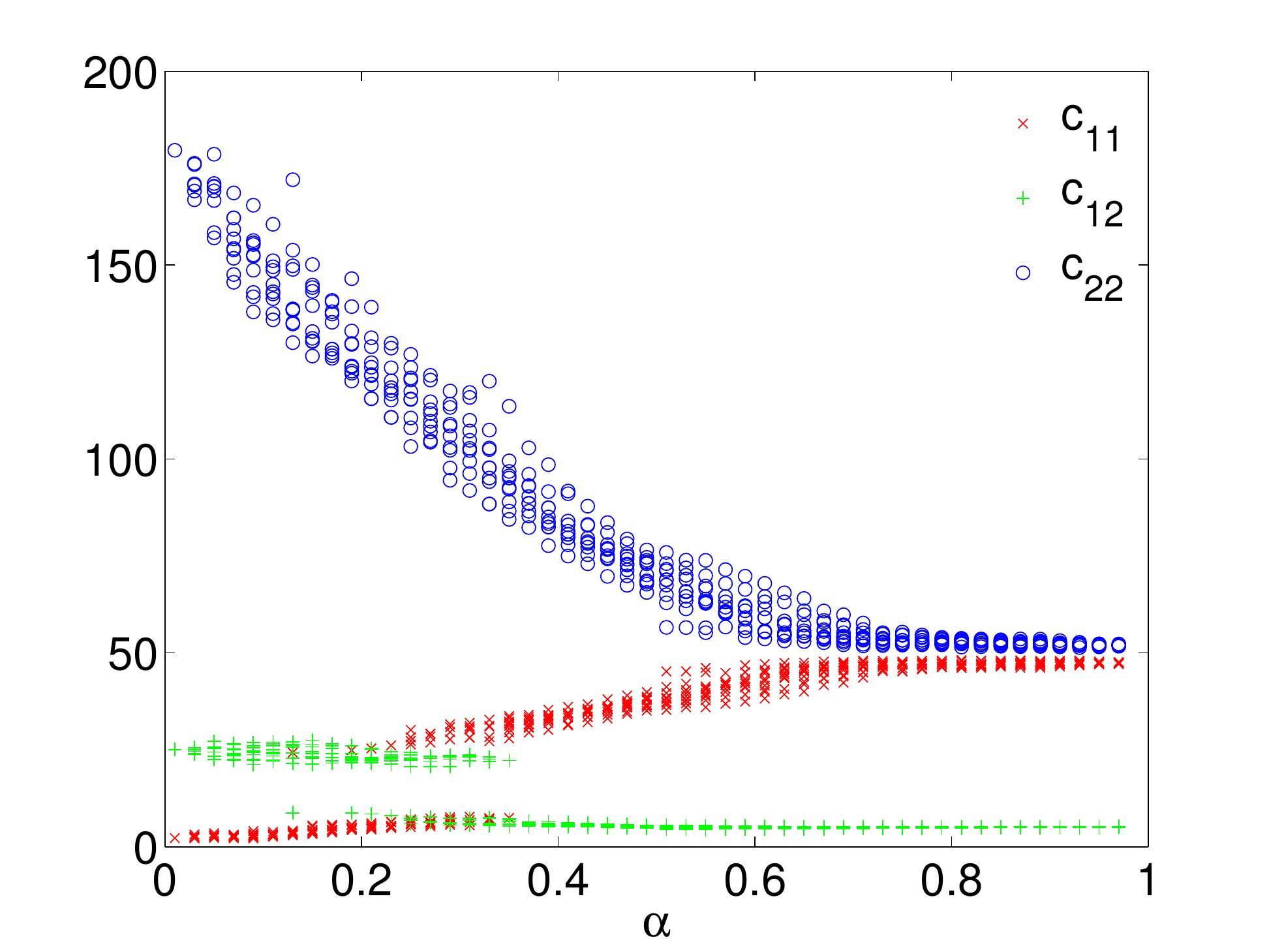} 
	\caption{Semisupervised learning in a network of political blogs~\cite{adamic2005political}.  Different points correspond to independent runs with different initial labels.  On the top left, the best possible parameters $q_a$ and $c_{ab}$ are given to the algorithm in advance.  On the top right, the algorithm learns these parameters using an EM algorithm, seeded by the known labels.  The bottom panels show how the learned $q_1$ and $c_{ab}$ change as $\alpha$ increases, moving from a core-periphery structure where nodes are divided according to high or low degree, to the correct assortative structure where they are divided along political lines.}\label{fig:polblogs}
\end{figure}

We test two real networks, namely a network of political blogs~\cite{adamic2005political} and Zachary's karate club network.  The blog network is composed of $1222$ blogs and links between them that were active during the 2004 US elections; human curators labeled each blog as liberal or conservative.  In Fig.~\ref{fig:polblogs} we plot the overlap between the inferred labels and the ground truth labels, with multiple independent runs of BP with different initial labels.  This network is known not to be well-modeled by the stochastic block model, since the highest-likelihood SBM splits the nodes into a core-periphery structure with high-degree nodes in the core and low-degree nodes outside it, instead of dividing the network along political lines~\cite{Karrer2011stochastic}.  Indeed, as the top left panel shows, even given the correct parameters, BP often falls into a core-periphery structure instead of the correct one.  However, once $\alpha$ is sufficiently large, we move into the basin of attraction of the correct division.  

On the top right of Fig.~\ref{fig:polblogs}, we see a similar transition, but now because the EM algorithm succeeds in learning the correct parameters.  There are two fixed points of the learning process in parameter space, corresponding to the high/low degree division and the political one.  Both of these are local minima of the free energy~\cite{Zhang2012comparative}.  As $\alpha$ increases, the correct one becomes the global minimum, and its basin of attraction gets larger, until the fraction of runs that arrive at the correct parameters (and therefore an accurate partition) becomes large.

We show this learning process in the lower panels of Fig.~\ref{fig:polblogs}.  Since there are just two groups, $q_1$ determines the group sizes, where we break symmetry by taking $q_1$ to be the smaller group.  As $\alpha$ increases, we move from $q_1=0.3$ to $q_1=0.47$.  On the lower right, we see the parameters $c_{ab}$ change from a core-periphery structure with $c_{22} > c_{12} > c_{11}$ to an assortative one with $c_{11} \approx c_{22} > c_{12}$.

\begin{figure}
   \centering
    \includegraphics[width=0.45\columnwidth]{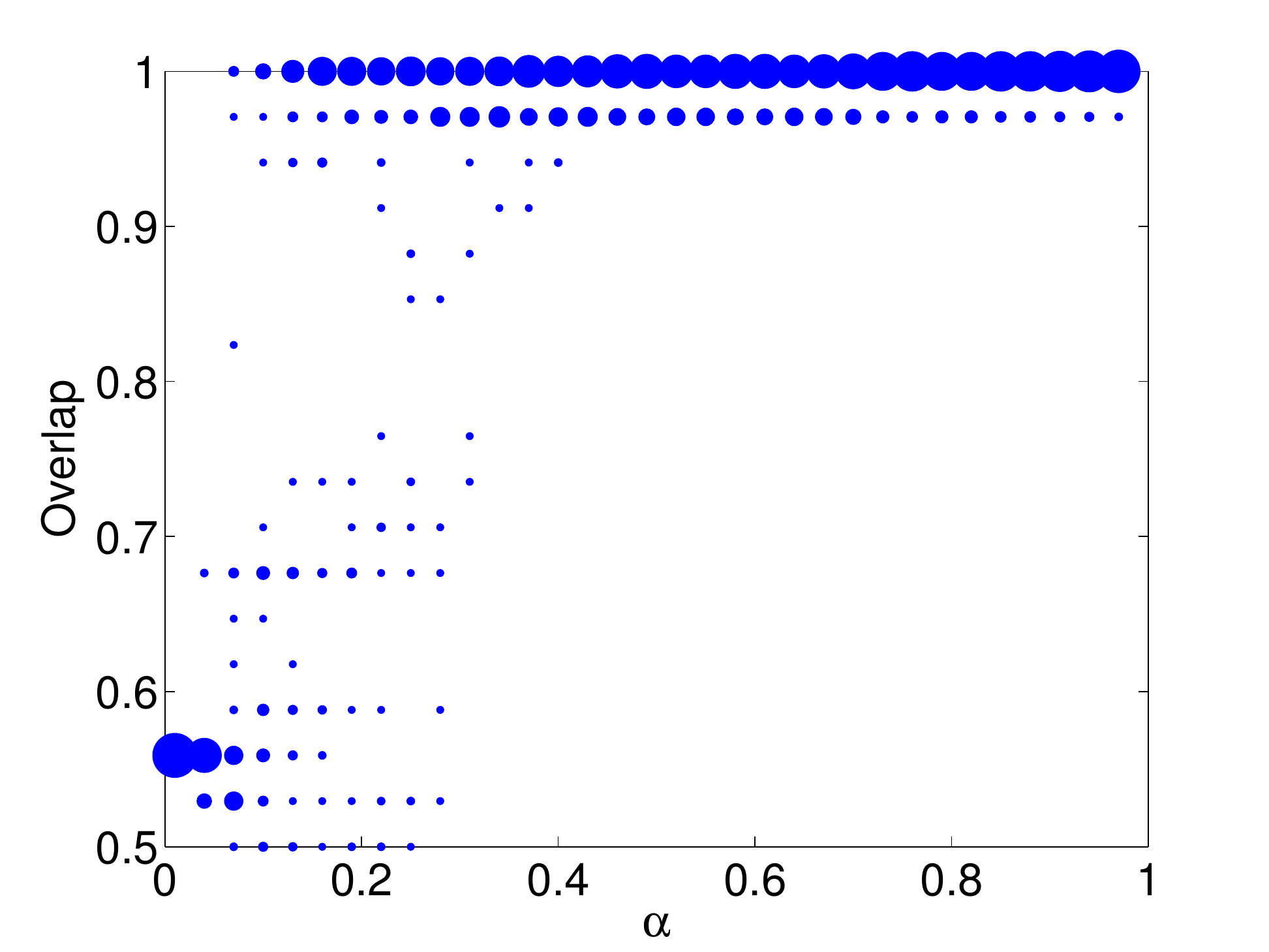} 
    \includegraphics[width=0.45\columnwidth]{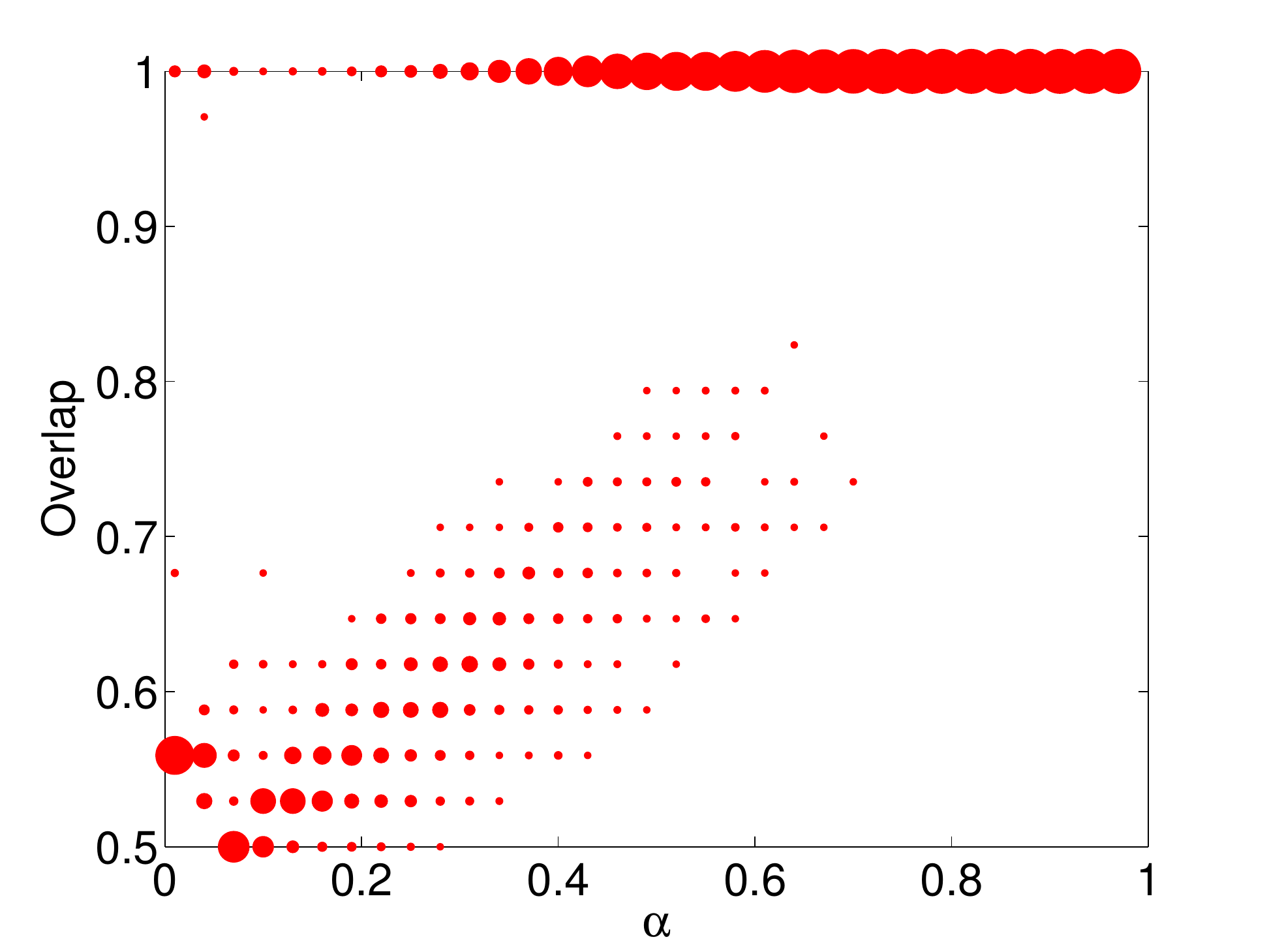} 
    \includegraphics[width=0.45\columnwidth]{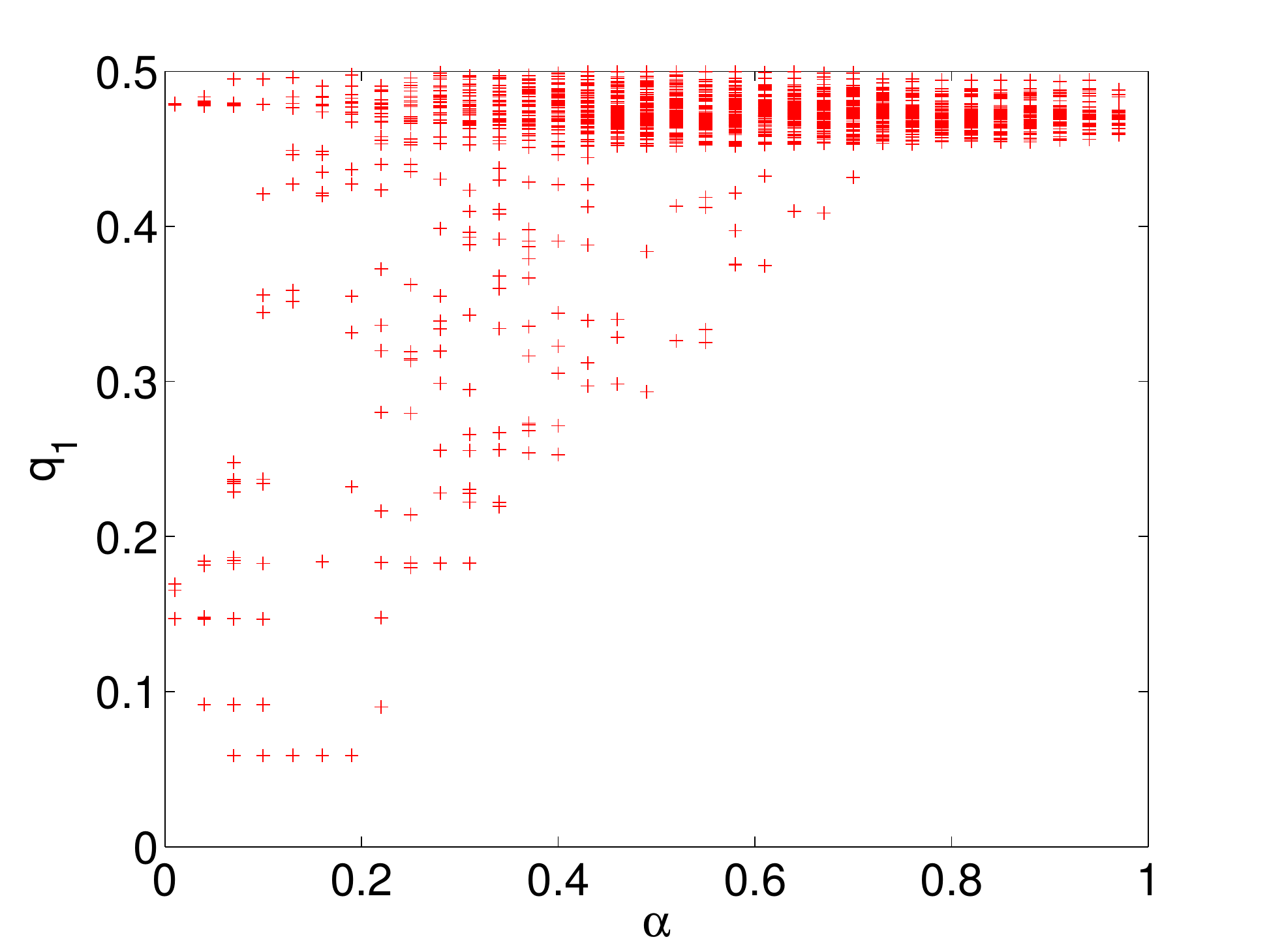} 
    \includegraphics[width=0.45\columnwidth]{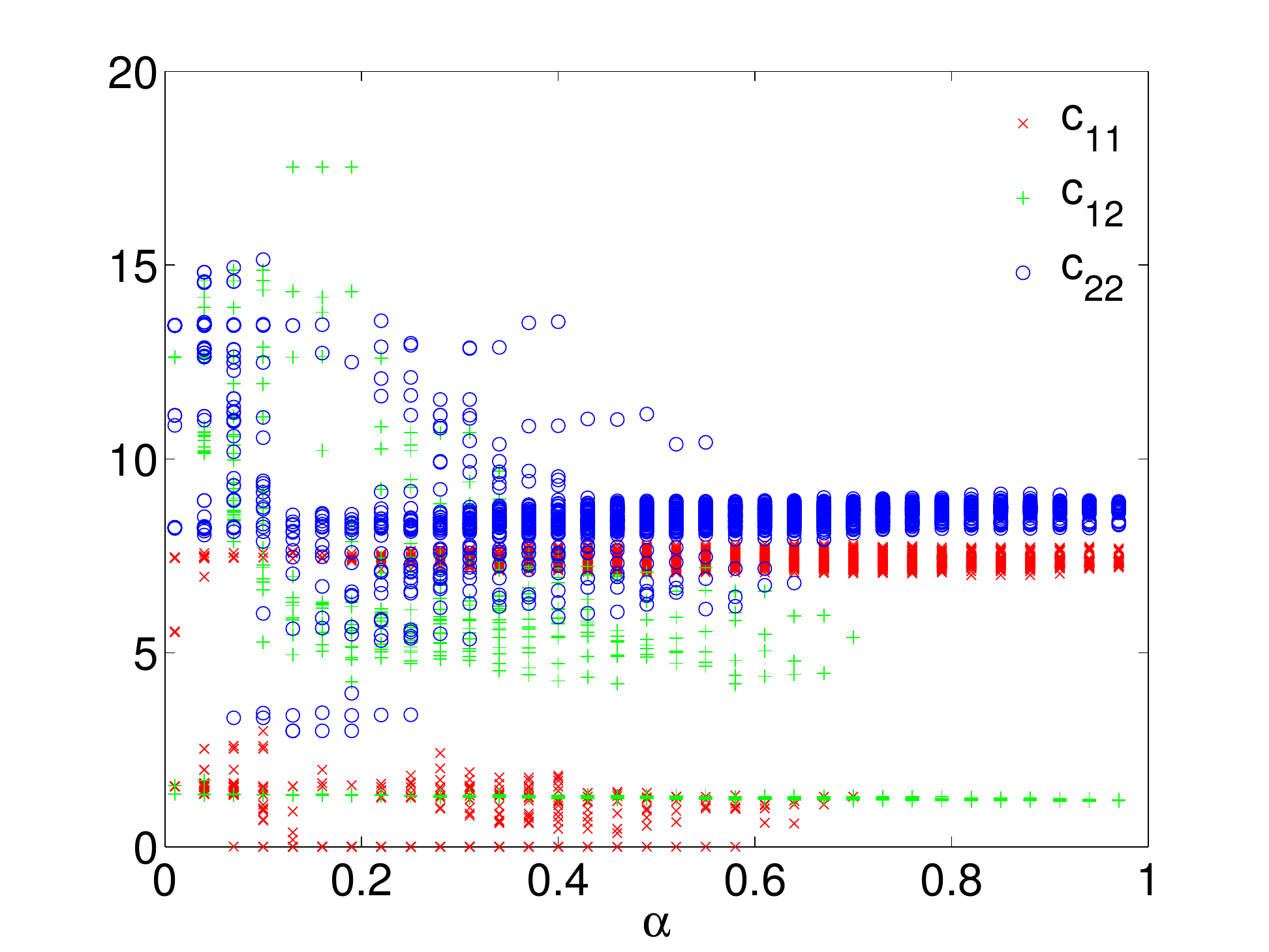} 
	\caption{Semisupervised learning in Zachary's karate club~\cite{zachary1977information}, with experiments analogous to Fig.~\ref{fig:polblogs}: in the upper left, the optimal parameters are given to the algorithm in advance, while in the upper right it learns them with an EM algorithm, giving the parameters shown in the bottom panels.  As in the political blog network, the algorithm makes a transition from a core-periphery structure to the correct assortative structure.}
	\label{fig:karate}
\end{figure}

Our second example is Zachary's Karate club~\cite{zachary1977information}, which represents friendship patterns between the 34 members of a university karate club which split into two factions.  As with the blog network, it has two local optima in parameter place, one corresponding to a high/low degree division (which in the unsupervised case has lower free energy) and the other into the two factions~\cite{Decelle2012asymptotic}.  We again do two types of experiments, one where the best parameters $q_a, c_{ab}$ known in advance, and the other where we learn these parameters with the EM algorithm.  Our results are showin in Fig~\ref{fig:karate} and are similar to Fig.~\ref{fig:polblogs}.  As $\alpha$ increases, the overlap improves, in the first case because the known labels push us into the basin of attraction of the correct division, and in the second case because the EM algorithm finds the correct parameters.  

As discussed elsewhere~\cite{Karrer2011stochastic,Decelle2012asymptotic}, the standard stochastic block model performs poorly on these networks in the unsupervised case.  It assumes a Poisson degree distribution within each community, and thus tends to divide nodes into groups according to their degree; in contrast, these networks (and many others) have heavy-tailed degree distributions within communities.  A better model for these networks is the degree-corrected stochastic block model~\cite{Karrer2011stochastic}, which achieves a large overlap on the blog network even when no labels are known.  We emphasize that our analysis can easily be carried out for the degree-corrected SBM as well, 
using the BP equations given in~\cite{yan2012model}.  On the other hand, it is interesting to observe how, in the semisupervised case, even the standard SBM succeeds in recognizing the network's structure at a moderate value of $\alpha$.

%


\section{Conclusion and discussion}
\label{sec:con}

We have studied semisupervised learning in sparse networks with belief propagation and the stochastic block model, focusing on how the detectability and easy/hard transitions depend on the fraction $\alpha$ of known nodes.  In agreement with previous work based on a zero-temperature approximation~\cite{allahverdyan2010community,steeg2013phase}, for $k=2$ groups the detectability transition disappears for $\alpha > 0$.  However, for large $k$ where there is a hard but detectable phase in the unsupervised case, the easy/hard transition persists up to a critical value of $\alpha$, creating a line of discontinuities in the overlap ending in a second-order phase transition.  

We found qualitatively similar behavior in two real networks, where the overlap jumps discontinuously at a critical value of $\alpha$.  When the best possible parameters of the block model are known in advance, this happens when the basin of attraction of the correct structure becomes larger; when we learn them with an EM algorithm as in~\cite{Decelle2011inference,Decelle2012asymptotic}, it occurs because the optimal parameters become global minima of the free energy.  In particular, even though the standard block model is not a good fit to networks like the blog network or the karate club, where each community has a heavy-tailed degree distributions, we found that at a certain value of $\alpha$ it switches from a core-periphery structure to the correct assortative structure.

It would be very interesting to apply this formalism to \emph{active} learning, where rather than learning the labels of a random set of $\alpha n$ nodes, the algorithm must choose which nodes to explore.  One approach to this problem~\cite{active-kdd} is to explore the node with the largest mutual information between it and the rest of the network, as estimated by Monte Carlo sampling of the Gibbs distribution, or (more efficiently) using belief propagation.  We leave this for future work.


\begin{acknowledgments}
C.M. and P.Z. were supported by AFOSR and DARPA under grant FA9550-12-1-0432.  We are grateful to Florent Krzakala, Elchanan Mossel, and Allan Sly for helpful conversations.
\end{acknowledgments}

\bibliography{zp}

\end{document}